\documentclass[aps,apl,twocolumn,groupedaddress ]{revtex4-1}
\usepackage{graphicx}
\usepackage{amsmath}
\usepackage{color}

\begin{document}


\title{Frequency-Tunable Microwave Field Detection in an Atomic Vapor Cell}


\author{Andrew Horsley}
\email[]{andrew.horsley@unibas.ch}
\author{Philipp Treutlein}
\email[]{philipp.treutlein@unibas.ch}
\affiliation{Departement Physik, Universit\"{a}t Basel, CH-4056 Switzerland}



\date{\today}

\begin{abstract}
We use an atomic vapor cell as a frequency tunable microwave field detector operating at frequencies from GHz to tens of GHz. We detect microwave magnetic fields from 2.3~GHz to 26.4~GHz, and measure the amplitude of the $\sigma_+$ component of an 18~GHz microwave field. Our proof-of-principle demonstration represents a four orders of magnitude extension of the frequency tunable range of atomic magnetometers from their previous dc to several MHz range. When integrated with a high-resolution microwave imaging system~\cite{Horsley2015}, this will allow for the complete reconstruction of the vector components of a microwave magnetic field and the relative phase between them. Potential applications include near-field characterisation of microwave circuitry and devices, and medical microwave sensing and imaging.
\end{abstract}


\maketitle

Atomic vapor cell magnetometers are among the most sensitive detectors for magnetic fields~\cite{Budker2007,Savukov2005b,Sheng2013}, and show promise in applications including gyroscopes~\cite{Donley2009}, explosives detection~\cite{Lee2006}, materials characterization~\cite{Romalis2011,Deans2016,Wickenbrock2016}, in MRI for both medical~\cite{Savukov2013a} and microfluidics applications~\cite{Xu2008}, and for magnetic imaging of the human heart~\cite{Bison2009,Alem2015} and brain~\cite{Savukov2013b,Johnson2013,Wyllie2012,Shah2013}. Tunable atomic magnetometers have previously operated from dc to rf frequencies of a few MHz. In this letter, we demonstrate the principle of a continuously frequency tunable microwave magnetometer, capable of detecting magnetic fields from GHz to tens of GHz frequencies. This represents a four orders of magnitude extension of the atomic magnetometer frequency tunable range. Potential applications include near-field characterisation of microwave circuitry~\cite{Sayil2005,Dubois2008,Sardi2015,Horsley2015}, and medical microwave sensing and imaging~\cite{Fear2002,Nikolova2011,Chandra2015}, where high resolution, intrinsically calibrated atomic sensors could replace bulky and field-perturbing antennas.

At the fixed microwave frequency of 6.8~GHz, we have previously used atoms to produce polarisation-resolved images of microwave magnetic near-field distributions~\cite{Boehi2010a,Boehi2012,Horsley2013,Affolderbach2015,Horsley2015}. We have achieved $50\times50\times140\,\mu\mathrm{m}^3$ spatial resolution and $1\,\mu\mathrm{T}/\mathrm{Hz}^{-1/2}$ sensitivity, observing good agreement between the measured fields and simulations for simple test structures~\cite{Horsley2015}. (Atom-like) NV centres have recently been used for nanoscale near-field detection~\cite{Wang2015} and imaging~\cite{Appel2015}, and microwave \emph{electric} field imaging has also been demonstrated at discrete frequencies using Rydberg atoms~\cite{Sedlacek2012,Holloway2014a,Fan2014,Fan2015}. To-date, however, there has been no atomic sensor available for the detection of fields at arbitrary microwave frequencies, as required for most applications.

Our frequency-tunable sensing scheme is able to operate in either of two modes: to detect microwaves of unknown frequency (i.e. as a microwave spectrum analyser); or to accurately measure the amplitudes of microwave magnetic fields (B$_{\mathrm{mw}}$) of known frequency. In this letter, we focus on the latter, measuring B$_{\mathrm{mw}}$ through coherent Rabi oscillations driven by the microwave on atomic hyperfine transitions~\cite{Boehi2010a,Boehi2012}. The atomic transitions are sensitive to a narrow microwave frequency band, and we use a dc magnetic field (B$_{\mathrm{dc}}$) to tune the transition frequency to the desired value. This is conceptually similar to rf magnetometry, where a dc magnetic field is used to tune the Zeeman splitting of adjacent m$_F$ states~\cite{Savukov2005b}. However, much larger dc fields are required for microwave frequencies, bringing us into the hyperfine Paschen-Back regime. Moreover, since our technique is based on measurements of frequency rather than amplitude, it is intrinsically calibrated and less sensitive to environmental noise. In this letter, we work with the $^{87}$Rb hyperfine ground states, however our technique is applicable to any microwave magnetic dipole transition with optical read-out of one of the states, such as other alkali atoms and NV centres.

\begin{figure}
\includegraphics[width=0.35\textwidth]{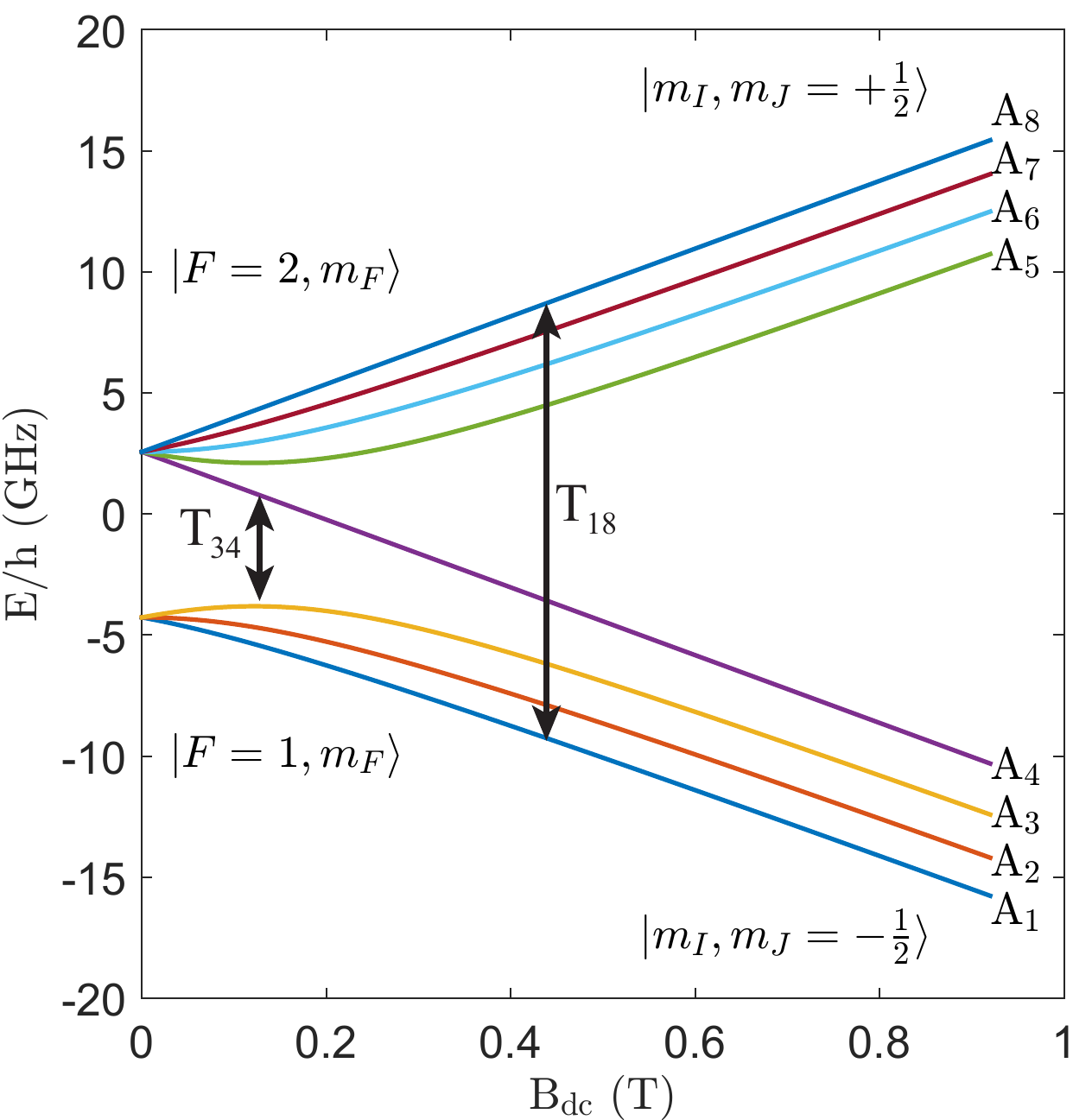}%
\caption{\label{fig:Rb87_HyperfineShift}(Color online) Energy splitting of the $^{87}$Rb hyperfine ground state levels as a function of applied dc magnetic field. Arrows indicate the hyperfine transitions used in this work. }
\end{figure}

Figure~\ref{fig:Rb87_HyperfineShift} shows the splitting of the $^{87}$Rb 5S$_{1/2}$ ground state in an applied dc magnetic field, calculated using the Breit-Rabi formula~\cite{Steck87}. We label the 8 hyperfine states A$_1 \rightarrow \mathrm{A}_8$, in order of increasing energy. At B$_{\mathrm{dc}}=0$, the states are grouped according to the $|F,m_F\rangle$ basis defined by the atomic Hamiltonian, and split by 6.8~GHz. As B$_{\mathrm{dc}}$ is increased, the hyperfine states move towards the $|I,m_I,J,m_J\rangle$ basis defined by the dc field. This is accompanied by a corresponding shift in transition frequencies and coupling strengths between the hyperfine states. An applied microwave field will drive Rabi oscillations on a given transition that has been tuned to resonance. The amplitude of the microwave vector component $B_{\gamma}$ is given by the measured Rabi oscillation frequency, $\Omega_{\mathrm{R}}$, the transition matrix element, and well-known fundamental constants. For a microwave resonant with transition T$_{if}$, we have
\begin{equation}
B_{\gamma} = |\alpha_{if}| \frac{\hbar}{\mu_B} |\Omega_{\mathrm{R}}|,
\end{equation}
where $\gamma=-,\pi,+$ is the polarisation of the transition, and
\begin{equation}
\alpha_{if}\equiv \frac{1}{ 2\langle A_f| J_{\gamma}|A_i\rangle}.
\end{equation}
The coupling factor $\alpha_{if}$ is a function of the applied dc magnetic field. As B$_{\mathrm{dc}}$ increases, $\alpha\rightarrow 1$ for the $\sigma_+$ polarised transitions, and $\alpha\rightarrow 0$ for the $\sigma_-$ and $\pi$ transitions. For $^{87}$Rb, we can find $\sigma_-$ ($\pi$) transitions where $|\alpha|>0.15$ for microwave frequencies within the range $2.2-9.1\,\mathrm{GHz}$ ($5.9-22.8\,\mathrm{GHz}$)~\cite{Supplementary}. Strong $\sigma_+$ transitions are available for all frequencies. Previous B$_{\mathrm{mw}}$ reconstruction schemes have relied on the presence of strong $\pi$ transitions~\cite{Horsley2015}. However, it is possible to completely reconstruct the vector components of the microwave magnetic field amplitude, and the relative phases between them, using only $\sigma_+$ transitions~\cite{Supplementary}. The reconstruction then requires measurements of B$_+$ with the dc magnetic field both parallel and anti-parallel to each of the lab-frame $x$, $y$, and $z$ axes. In this letter, we address two microwave transitions: the $\sigma_+$ polarised T$_{18}$ transition, and the $\sigma_-$ polarised T$_{34}$ transition.

\begin{figure}
\includegraphics[width=0.5\textwidth]{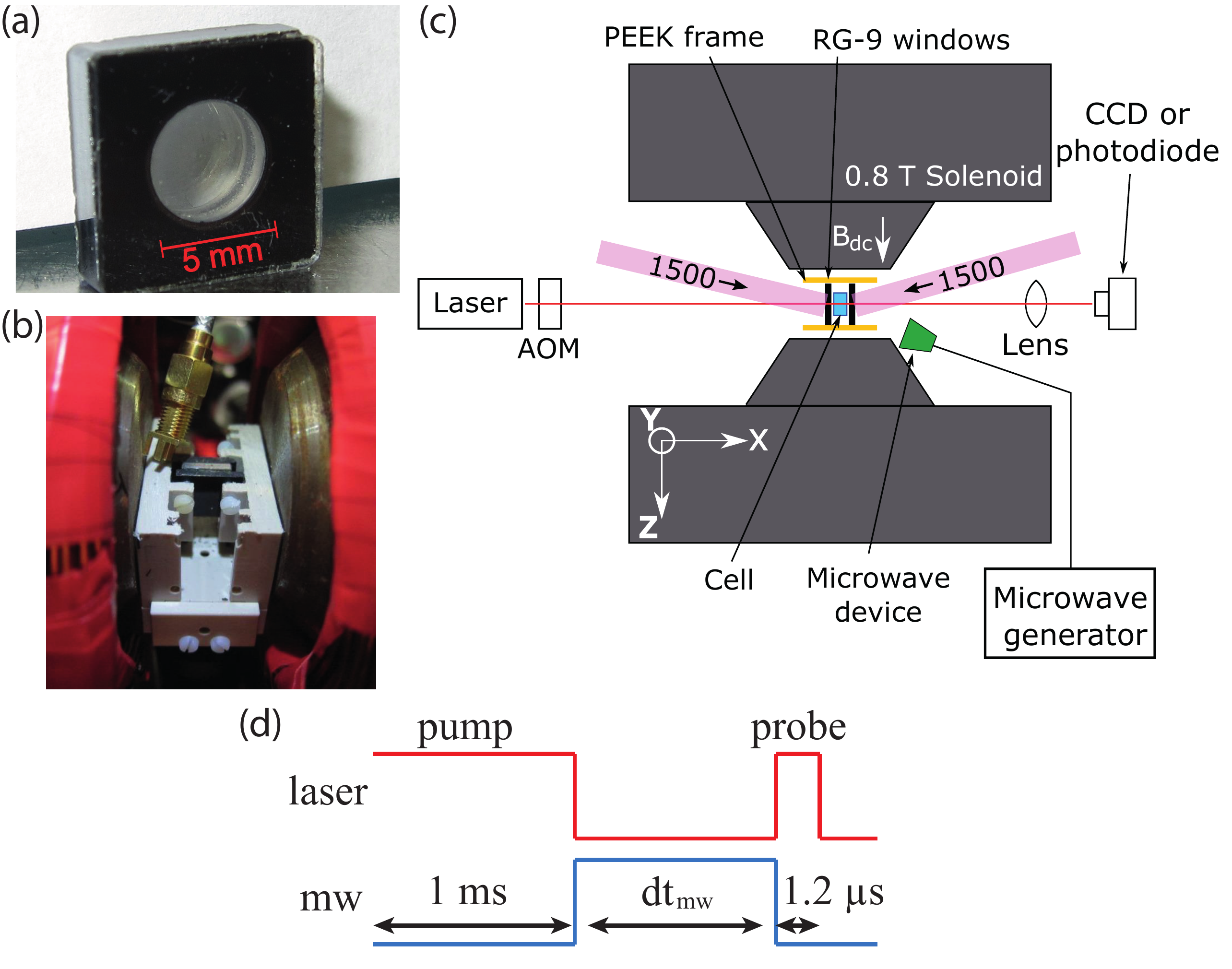}%
\caption{\label{fig:FreqTunSetup}(Color online) (a) The vapor cell; (b) the experiment setup, showing the vapor cell sandwiched between two RG-9 glass pieces and mounted on the PEEK frame inside the solenoid. The gold microwave output coupler can be seen above the cell; (c) schematic of the experiment setup; (d) the Rabi sequence used to obtain microwave amplitudes. }
\end{figure}

We use a 2~mm thick, microfabricated vapor cell (Fig.~\ref{fig:FreqTunSetup}(a)), filled with natural Rb and 63~mbar of N$_2$ buffer gas~\cite{Pellaton2012,Horsley2013}. Doppler and collisional broadening result in an optical linewidth of 1.5~GHz. We use a 780~nm laser, linearly polarised along B$_{\mathrm{dc}}$, for both optical pumping and probing. A water-cooled solenoid~\footnote{Brukner, B-E 10 solenoid with B-MN C4 power supply, built in 1974.} provides dc magnetic fields up to 0.8~T (Fig.~\ref{fig:FreqTunSetup}.(b)). We heat the vapor cell using a 2~W laser at  1500~nm~\cite{Mhaskar2012,Supplementary}, and obtain the cell temperature by fitting optical absorption spectra (taken at B$_{\mathrm{dc}}=0$) with the model described in Ref.~\cite{Zentile2015}.

\begin{figure}
\includegraphics[width=0.5\textwidth]{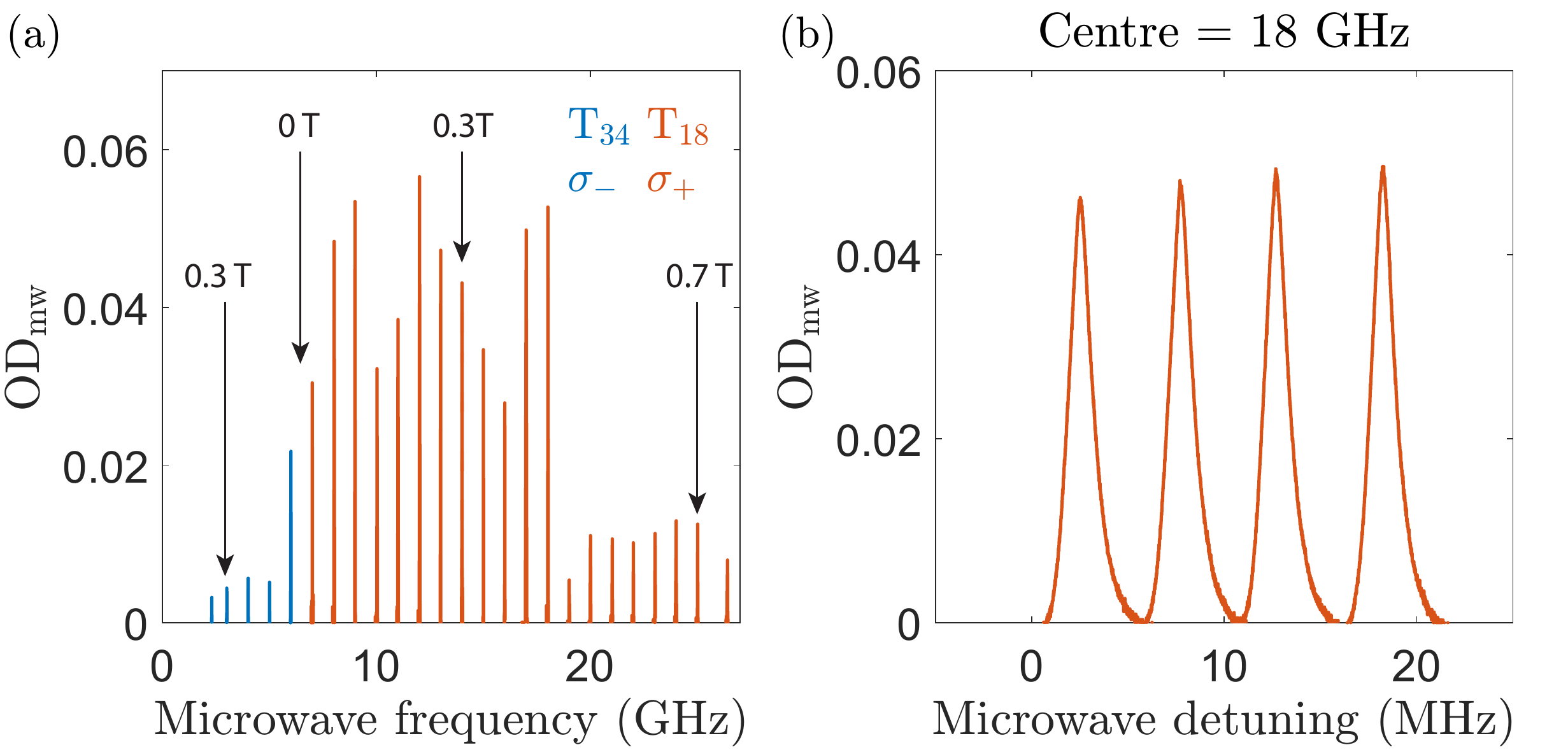}%
\caption{\label{fig:DR_freqtun_paper}(Color online) Detection of microwave signals at arbitrary frequencies. (a) DR detection of frequencies ranging from 2.3~GHz to 26.4~GHz. Each peak corresponds to a single experimental run taken using a different B$_{\mathrm{dc}}$. The data was taken on both the T$_{34}$ (blue) and T$_{18}$ (red) transitions, and shows the change in optical density induced by the microwave. Labels in black indicate relevant B$_{\mathrm{dc}}$ strengths. (b) DR peaks for frequencies near 18~GHz, showing fine-tuning of the detection frequency by adjusting B$_{\mathrm{dc}}$.}
\end{figure}

\begin{figure}[t!]
\includegraphics[width=0.5\textwidth]{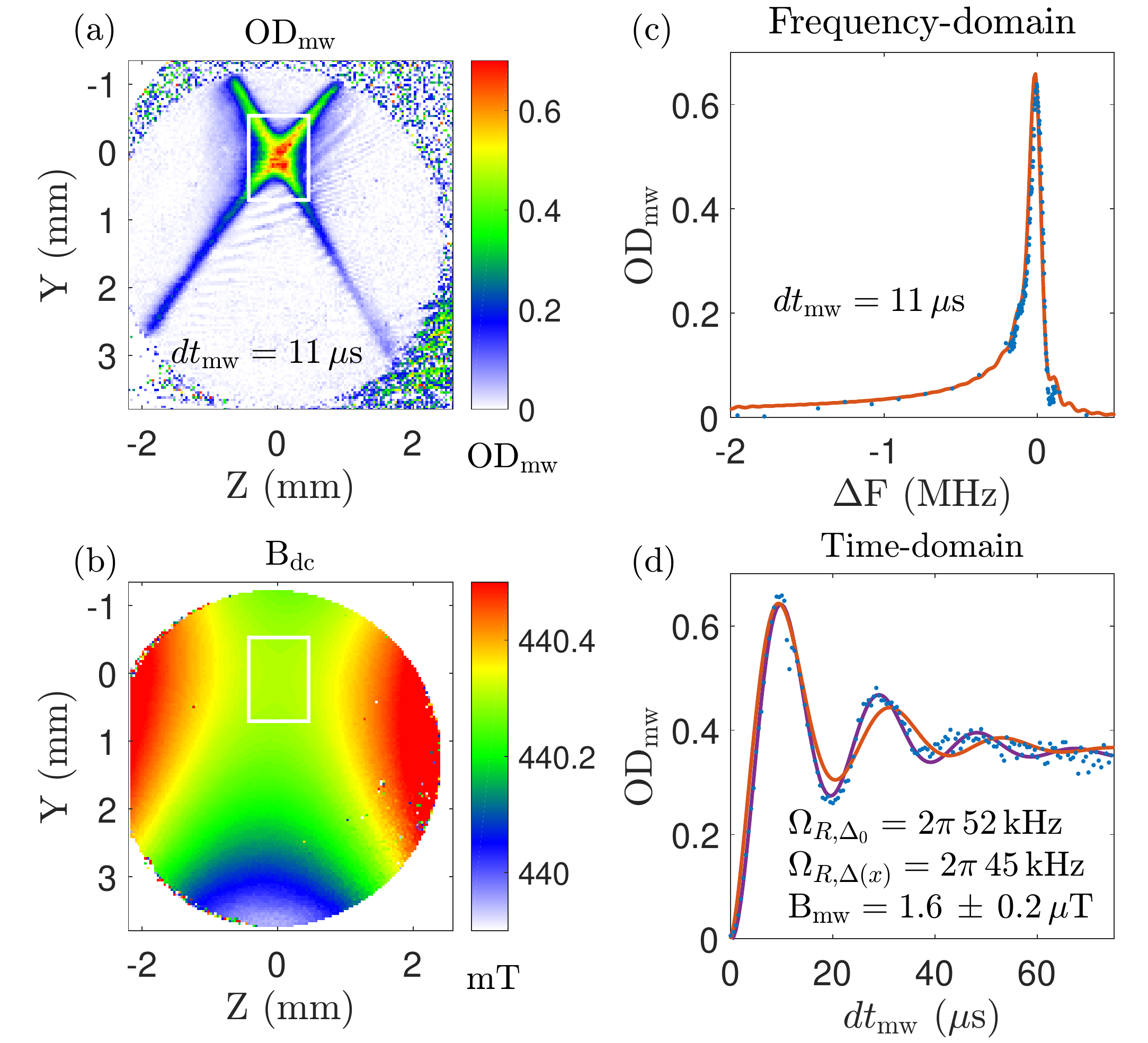}%
\caption{\label{fig:Bmw_Bdc_18GHz}(Color online) Measuring an 18.003~GHz microwave field amplitude. Experimentally obtained images of (a)  OD$_{\mathrm{mw}}$ for a microwave pulse resonant with the T$_{18}$ transition in the B$_{\mathrm{dc}}$ field centre; (b) map of the local B$_{\mathrm{dc}}$ field extracted from measurements as in (a). The circular field of view is given by the cell walls, and the white boxes indicate the region appropriate for microwave sensing; (c+d) scans in the frequency and time domains, for the pixel at ($z=2.29$~mm, $y=1.32$~mm). Fits are shown assuming a constant microwave detuning along the $x$ axis (purple, time domain only, $\Omega_{\mathrm{R},\Delta_0}$), and allowing for an inhomogeneous detuning (red, simultaneous fit to time and frequency domain data,  $\Omega_{\mathrm{R},\Delta(x)}$).}
\end{figure}

We can detect microwaves over a broad range of frequencies. Figure~\ref{fig:DR_freqtun_paper}(a) shows experimentally obtained double-resonance (DR) peaks for microwaves spanning frequencies from 2.3~GHz to 26.4~GHz. The laser beam is applied continuously, and its frequency is tuned to resonance with one of the hyperfine states. This partially depopulates the state through optical pumping. Through the transmission of the laser onto a photodiode, we monitor the change in optical density (OD$_{\mathrm{mw}}$) as we sweep the frequency of an applied microwave field. Whenever the microwave frequency crosses a hyperfine transition coupling to the optically pumped state, we see a peak in OD$_{\mathrm{mw}}$, corresponding to repopulation of the state. Each peak in Fig.~\ref{fig:DR_freqtun_paper} represents a separate measurement, with B$_{\mathrm{dc}}$ changed between measurements. Peaks in the range $7-26.4$~GHz (red) were measured on the $\sigma_+$ T$_{18}$ transition, whilst peaks in the range $2.3-6$~GHz (blue) were measured on the $\sigma_-$ T$_{34}$ transition. The laser beam probed the approximate spatial centre of the dc magnetic field (see Fig.~\ref{fig:Bmw_Bdc_18GHz}(b)), with inhomogeneities in the dc field resulting in a 1~MHz (FWHM) DR linewidth. The peak area is proportional to $\Omega_{R}$, however it is also sensitive to fluctuations in the laser intensity and cell temperature, and calibration with a known field is required to extract $\Omega_{R}$. For fixed conditions (not shown), we observe the expected linear scaling of the peak area with the square-root of the microwave power. The lower limit on microwave frequency in Fig.~\ref{fig:DR_freqtun_paper}(a) was imposed by the asymptotic behaviour of the T$_{34}$ transition frequency at high B$_{\mathrm{dc}}$~\cite{Supplementary}. Other transitions (e.g. T$_{45}$) are available with frequencies down to dc, in which case the lower detection limit will be given by the optical distinguishability of neighbouring hyperfine states, in general on the order of the 0.5~GHz Doppler broadening in a cell without buffer gas. The upper frequency limit is given only by technical limitations, e.g. the available B$_{\mathrm{dc}}$ strength, or in our case the microwave frequency generator.

The dc magnetic field provides fine control over the microwave detection frequency, as shown in Fig.~\ref{fig:DR_freqtun_paper}(b). Each DR peak represents a single measurement on the T$_{18}$ transition, with the zero of the horizontal axis corresponding to a microwave frequency of 18~GHz. By scanning B$_{\mathrm{dc}}$, such DR measurements could be used in a spectrum analyser mode, to search for microwave fields of unknown frequency.

For an accurate measure of the B$_{\mathrm{mw}}$ amplitude, we need to measure the Rabi frequency directly. We demonstrate this by measuring the $\sigma_+$ amplitude of an 18~GHz microwave. The observed oscillations are a function of both microwave amplitude ($\propto \Omega_R$) and detuning from resonance ($\Delta$), with a frequency $\Omega=\sqrt{\Omega_R^2+\Delta^2}$. In general, it is simple to directly measure the detuning, e.g. from the peak of a DR scan. We use a relatively thick cell in our proof-of-principle setup, however, and must account for a non-negligible B$_{\mathrm{dc}}$ inhomogeneity along the $x$ axis, which the laser integrates over as it passes through the cell. Our experiment `Rabi' sequence, shown schematically in Fig.~\ref{fig:FreqTunSetup}(d), begins with an optical pumping pulse to depopulate the A$_1$ state. A microwave pulse of duration $dt_{\mathrm{mw}}$ then drives Rabi oscillations between the A$_1$ and A$_8$ states, and we detect the repopulation of the A$_1$ state (i.e. OD$_{\mathrm{mw}}$) with a $1.2\,\mu$s probe laser pulse. We repeat this sequence, scanning either the frequency (Fig.~\ref{fig:Bmw_Bdc_18GHz}(c)) or duration (Fig.~\ref{fig:Bmw_Bdc_18GHz}(d)) of the microwave pulse. The frequency domain scan primarily constrains B$_{\mathrm{dc}}$ (Fig.~\ref{fig:Bmw_Bdc_18GHz}(b)), and we extract the B$_{\mathrm{mw}}$ amplitude by simultaneously fitting the two scans~\cite{Supplementary}. In future setups, the frequency domain scan will be unnecessary.

We used the T$_{18}$ transition for sensing, with the frequency of the laser tuned to address the A$_1$ level. A solenoid field of B$_{\mathrm{dc}}\approx 0.44\,\mathrm{T}$ tuned the T$_{18}$ transition of atoms in the B$_{\mathrm{dc}}$ field centre to resonance with the applied 18~GHz microwave. At this dc field, the A$_1$ level is optically well-resolved~\cite{Horsley2015thesis}, and almost all of the measured optical density of $\mathrm{OD}=2.1$ was due to atoms in the A$_1$ state. The frequency domain scan was performed around 18.003~GHz with a fixed microwave pulse duration of $dt_{\mathrm{mw}}=11\,\mu$s, corresponding roughly to a $\pi$ pulse in the centre of the B$_{\mathrm{dc}}$ field, while the time domain scan was performed with a fixed frequency of 18.003~GHz.

We are working towards frequency tunable microwave imaging, and therefore use a CCD camera and the technique of absorption imaging for detection~\cite{Horsley2013a,Horsley2015thesis,Horsley2015}. The laser illuminates the entire cell, and we obtain images of OD$_{\mathrm{mw}}$ (Fig.~\ref{fig:Bmw_Bdc_18GHz}(a)). Due to inhomogeneities in the B$_{\mathrm{dc}}$ field, we only present microwave detection for a single pixel near the B$_{\mathrm{dc}}$ centre. If only single-channel sensing is required, real-time monitoring of the Rabi oscillations should also be possible using a photodiode and an off-resonant probe laser. The atomic response traced a contour line of the B$_{\mathrm{dc}}$ field in the cell, as illustrated in Fig.~\ref{fig:Bmw_Bdc_18GHz}(a), which shows an image of OD$_{\mathrm{mw}}$ for an $11\,\mu$s long 18.003~GHz microwave pulse. Scanning the microwave frequency (Fig.~\ref{fig:Bmw_Bdc_18GHz}(c)), we could quickly estimate the local T$_{18}$ transition frequency from the peak in OD$_{\mathrm{mw}}$. Using the Breit-Rabi formula, this transition frequency yielded an image of the applied B$_{\mathrm{dc}}$, shown in Fig.~\ref{fig:Bmw_Bdc_18GHz}(b). We see that the solenoid field is saddle-shaped, homogeneous to better than $10^{-3}$, but nonetheless varying by several Gauss over the cell. The plane of the image is slightly offset from the $x=0$ centre of the solenoid field.

The simultaneous fit of the Fig.~\ref{fig:Bmw_Bdc_18GHz}(c+d) frequency and time domain data, described in detail in the supplementary material, yielded a Rabi frequency of $\Omega_{\mathrm{R},\Delta(x)}=2\pi\,45\,$kHz, corresponding to B$_{\mathrm{mw}}=1.6\,\mu$T. The shape of the frequency scan data in Fig.~\ref{fig:Bmw_Bdc_18GHz}(c) is primarily determined by the variation in B$_{\mathrm{dc}}$ along the $x$ axis, and is well-matched by the fit. The peak corresponds to atoms around the B$_{\mathrm{dc}}$ centre, where the field is relatively flat, while the long tail is due to the drop-off of B$_{\mathrm{dc}}$ away from the centre. The time domain fit in Fig.~\ref{fig:Bmw_Bdc_18GHz}(d) underestimates the oscillation frequency, due to constraints imposed by the simultaneous fitting of time and frequency domain data with our relatively simple model. In fact, we find much better agreement with the time domain data if we simply assume a constant $\Delta_0=-4\,\mathrm{kHz}$, given by the frequency domain peak position, and fit to the time domain data only. This is shown by the purple line in Fig.~\ref{fig:Bmw_Bdc_18GHz}(d), and we obtain $\Omega_{\mathrm{R},\Delta_0}=2\pi\,52\,$kHz, corresponding to B$_{\mathrm{mw}}=1.8\,\mu$T. By neglecting the range of larger detunings along the $x$ axis, this value serves as an upper bound on B$_{\mathrm{mw}}$. We can therefore estimate a B$_{\mathrm{mw}}$ accuracy of $10\%$ in our current proof-of-principle setup. We emphasise that in a high resolution setup using a thinner vapor cell~\cite{Horsley2015}, the image analysis will be much simpler, with B$_{\mathrm{mw}}$ directly extracted from time domain data and accuracy on the $10^{-3}$ level.

In conclusion, we have presented a proof-of-principle demonstration of microwave magnetic field detection for microwaves of any frequency, obtaining the amplitude of the $\sigma_+$ component of an 18~GHz microwave field, and detecting magnetic fields at frequencies from 2.3~GHz to 26.4~GHz. Magnetic field detection of frequencies up to 50~GHz should be achievable with room-temperature iron-bore solenoids (which can operate to $\sim$1.6~T), and given sufficient dc field homogeneity, it may be possible to detect fields up to 1~THz using the strongest available superconducting solenoids ($>$35~T). It would be interesting to explore the frequency tunability presented here with other detection techniques, such as Faraday rotation or EIT, and with other systems, such as NV centres~\cite{Appel2015}.

Our frequency tunable detection provides an essential step towards an atom-based characterisation tool for microwave devices operating at any frequency. To realise this in practical form, the frequency tunability demonstrated in this letter should now be integrated with a high resolution imaging setup~\cite{Horsley2015}. We note, however, that for some devices, such as circulators, the large applied B$_{\mathrm{dc}}$ fields may significantly perturb the device operation. High spatial resolution, provided by a thin vapor cell, will have the added benefit of significantly reducing our sensitivity to B$_{\mathrm{dc}}$ inhomogeneities. For a $200\,\mu$m thick cell operating in the B$_{\mathrm{dc}}$ field shown in Fig~\ref{fig:Bmw_Bdc_18GHz}(b), B$_{\mathrm{dc}}$ variation along the $x$ axis should be negligible, on the order of $2\pi\times1$~kHz. The transverse variation over a $50\times50\,\mu\mathrm{m}^2$ pixel at the B$_{\mathrm{dc}}$ centre would be $2\pi\times0.2$~kHz, increasing to $2\pi\times80$~kHz at a distance 1~mm either side of the centre. This is entirely acceptable compared to the tens to hundreds of kHz Rabi frequencies observed in Ref.~\cite{Horsley2015}. Moreover, the B$_{\mathrm{dc}}$ inhomogeneity could also be reduced by using shimming coils. We expect the sensitivity to be slightly better than for the fixed-frequency imaging reported in Ref.~\cite{Horsley2015}, due to the suppression of Rb spin-exchange relaxation in large B$_{\mathrm{dc}}$ fields~\cite{Kadlecek2001}. The integrated setup could be miniaturised by replacing the bulky solenoid with permanent rare-earth magnets, with B$_{\mathrm{dc}}$ adjusted through the magnet separation~\cite{Weller2012c,Zentile2014}.


\begin{acknowledgments}
This work was supported by the Swiss National Science Foundation (SNFS). We thank G. Mileti, C. Affolderbach, and Y. P\'{e}tremand for providing the vapor cell, and the group of P. Maletinsky for the loan of microwave circuitry. We also thank M. Belloni for interesting discussions, and C. Affolderbach for careful reading of the manuscript.
\end{acknowledgments}

\bibliography{bibliography_frequency_tunable}

\end{document}



\title{Frequency-Tunable Microwave Field Detection in an Atomic Vapor Cell: Supplementary Information}


\author{Andrew Horsley}
\email[]{andrew.horsley@unibas.ch}
\author{Philipp Treutlein}
\email[]{philipp.treutlein@unibas.ch}
\affiliation{Departement Physik, Universit\"{a}t Basel, CH-4056 Switzerland}



\date{\today}



\maketitle

\section{Microwave Amplitude Extraction}

To account for the $x$ axis variation in B$_{\mathrm{dc}}$ that the laser integrates over as it passes through the vapor cell, we take data in both the time and frequency domains, and perform a simultaneous fit to both data sets. Broadly speaking, the frequency domain data constrains the B$_{\mathrm{dc}}$ variation, and the time domain data gives us the microwave Rabi frequency. In future high resolution setups, the vapor cell will be sufficiently thin that B$_{\mathrm{dc}}$ variation through the cell will be negligible, rendering this frequency domain scan and simultaneous fitting process unnecessary. For our current proof-of-principle setup, we fit the data using
\begin{align}\label{eq:fitting}
\mathrm{OD}_{\mathrm{mw}} =  \int dx \Big[A\frac{\Omega_R^2}{\Omega_R^2 + \Delta(x)^2} \sin^2\Big(\tfrac{1}{2}\sqrt{\Omega_R^2 + \Delta(x)^2} \, dt_{\mathrm{mw}}\Big) \\
\times\exp(-dt_{\mathrm{mw}}/\tau_2)\nonumber + B \frac{\Omega_R^2}{\Omega_R^2 + \Delta(x)^2} \Big(1-\exp(-dt_{\mathrm{mw}}/\tau_1)\Big) \Big],
\end{align}
where $A$, $B$, $\Omega_R$, $\Delta(x)$, $\tau_1$, and $\tau_2$ are fit parameters. The first term describes Rabi oscillations with a coherence time $\tau_2$. The second term is phenomenological, accounting for the diffusion of atoms from neighbouring regions of the cell, with a time constant $\tau_1$. We neglect variation in $\Omega_R$ along the $x$ axis, as there was little variation in distance from the microwave source for a given $y$-$z$ position through the cell. We include the B$_{\mathrm{dc}}$ inhomogeneity along the $x$ axis by using $\Delta(x) = \delta \omega_0 - \delta \omega_2 (x-x_0)^2 - \delta \omega_4 (x-x_0)^4$, where $x_0$ is the $x$ axis centre of the B$_{\mathrm{dc}}$ field. This form is derived from the measured B$_{\mathrm{dc}}$ variation along the $y$ axis (Fig.~4(b) in the main text), justified by the cylindrical symmetry of the solenoid.

For the pixel at ($z=0$~mm, $y=0$~mm), shown in Figs.~4(c+d) of the main text, the fit parameters are $A=2.1$, $B=0.88$, $\tau_1=15\,\mu$s, $\tau_2=18\,\mu$s, $x_0= 0.55\,\mathrm{mm}$, $\delta \omega_0=2\pi\,4.4\,$kHz, $\delta \omega_2 = 2\pi\,260\,\mathrm{kHz}/\mathrm{mm}^2$, $\delta \omega_4= 2\pi\,250\,\mathrm{kHz}/\mathrm{mm}^4$, and $\Omega_{\mathrm{R},\Delta(x)}=2\pi\,45\,$kHz. The coupling constant between the Rabi frequency and microwave amplitude is essentially a constant value of $\alpha_{18}=0.492$ over the cell. The extracted Rabi frequency of $\Omega_{\mathrm{R},\Delta(x)}=2\pi\,45\,$kHz thus corresponds to B$_{\mathrm{mw}}=1.6\,\mu$T. The coherence time is around $10\%$ of the optically pumped population lifetime (T$_1$), with the dominant dephasing mechanism caused by fluctuations in Zeeman shift due to atomic motion through the B$_{\mathrm{dc}}$ inhomogeneities.

\section{Further B$_{\mathrm{mw}}$ Detection Details}



Due to the large dc magnetic fields and spatial constraints imposed by the solenoid, we avoided using resistive heaters to control the cell temperature. Instead, the cell was placed between two 2~mm thick pieces of RG9 glass (strongly absorptive at 1500~nm, with better than $90\%$ transmission at 780~nm), and we heated the glass using a 2~W laser at the wavelength of 1500~nm. The direct heating of the cell windows ensures that there is minimal build-up of Rb on the windows, and the localised cell heating is advantageous for detecting microwave fields produced by temperature-sensitive devices. We were able to operate at relatively high temperatures with minimal reduction in T$_1$ lifetime, due to the suppression of Rb-Rb spin exchange relaxation in large dc magnetic fields~\cite{Kadlecek2001}.

For the data presented in Fig.~3 of the main text, the 780~nm laser intensity was $5\,\mathrm{mW}/\mathrm{cm}^2$ and the beam diameter was 0.6~mm. The data in Fig.~3(b) was taken at a cell temperature of $99^{\circ}$C.

For the data presented in Fig.~4 of the main text, the 780~nm laser beam was expanded to cover the entire vapor cell, with an intensity of $120\,\mathrm{mW}/\mathrm{cm}^2$ averaged over the central 3~mm. The cell temperature was $115^{\circ}$C. We repeated the frequency and time domain scans 3 and 5 times, respectively. Fitting was performed on the averaged data, however before further analysis, we binned the CCD pixels into $3\times3$ blocks to create $35.5\times35.5\,\mu\mathrm{m}^2$ image pixels. This reduced the computational intensity of the data analysis, with minimal reduction in image quality, as the image pixel size was still significantly smaller than the $0.11\,\mathrm{mm}$ diffusion distance of atoms during their coherence lifetime (for $P_{\mathrm{fill}}=63\,\mathrm{mbar}$ N$_2$, $T_{\mathrm{fill}}=80^{\circ}$C, $T_{\mathrm{cell}}=115^{\circ}$C, we have $D=3.4\,\mathrm{cm}^2/\mathrm{s}$. Using $\Delta x=\sqrt{2 D\,dt}$ and $dt=18\,\mu$s, $\Delta x=0.11\,\mathrm{mm}$). The frequency domain data was taken less than 5 minutes after the time domain data, allowing us to minimise the effects of drifts in the B$_{\mathrm{dc}}$ field, which occurred with a timescale of several minutes to tens of minutes. These drifts also complicated the practical implementation of shimming fields to homogenise the solenoid field. However, we note that no attempt was made to actively stabilise the solenoid, and that solenoid stability and homogeneity have likely advanced in the decades since the solenoid was purchased in 1974.

The metallic poles of the magnet will impose boundary conditions on the microwaves. This may be a significant perturbation for the detection of far-field microwaves, as the 26~mm separation of our solenoid poles is comparable to microwave wavelengths. However, we do not expect the poles to be a problem in the detection of microwave near-fields hundreds of micrometers to a few millimeters above a source, which would be the operation mode for microwave device characterisation as in Ref.~\cite{Horsley2015}.

Regarding the variation of the B$_{\mathrm{dc}}$ field along the $x$ axis, we can define three regions: a `good' central region, where the B$_{\mathrm{dc}}$ field is relatively flat and atoms are near resonance with the microwave; two `bad' middle regions, in front of and behind the `good' region, where the detuning is small enough for the atoms to still undergo Rabi oscillations, but at a significantly different oscillation frequency to the `good' region, thus acting to wash out the `good' signal; and two `neutral' outer regions, where atoms are so far detuned from the microwave that they do not interact. We minimised the `bad' effects by placing the vapor cell mount on a translation stage, and adjusting the cell position to minimise the double-resonance linewidth. In doing so, we aligned the front of the cell with the edge of the `good' region, meaning that the atoms only saw one `bad' region, and an extended `neutral' region. The sensing region for B$_{\mathrm{mw}}$ was primarily in the `good' region, thinner than the full 2~mm thickness of the cell.



\section{Hyperfine Transitions in an Arbitrary DC Magnetic Field}\label{sec:hyperfine_trasitions_arbitrary_Bdc}

\begin{figure}[t!]
\centering
\includegraphics[width=8.7cm]{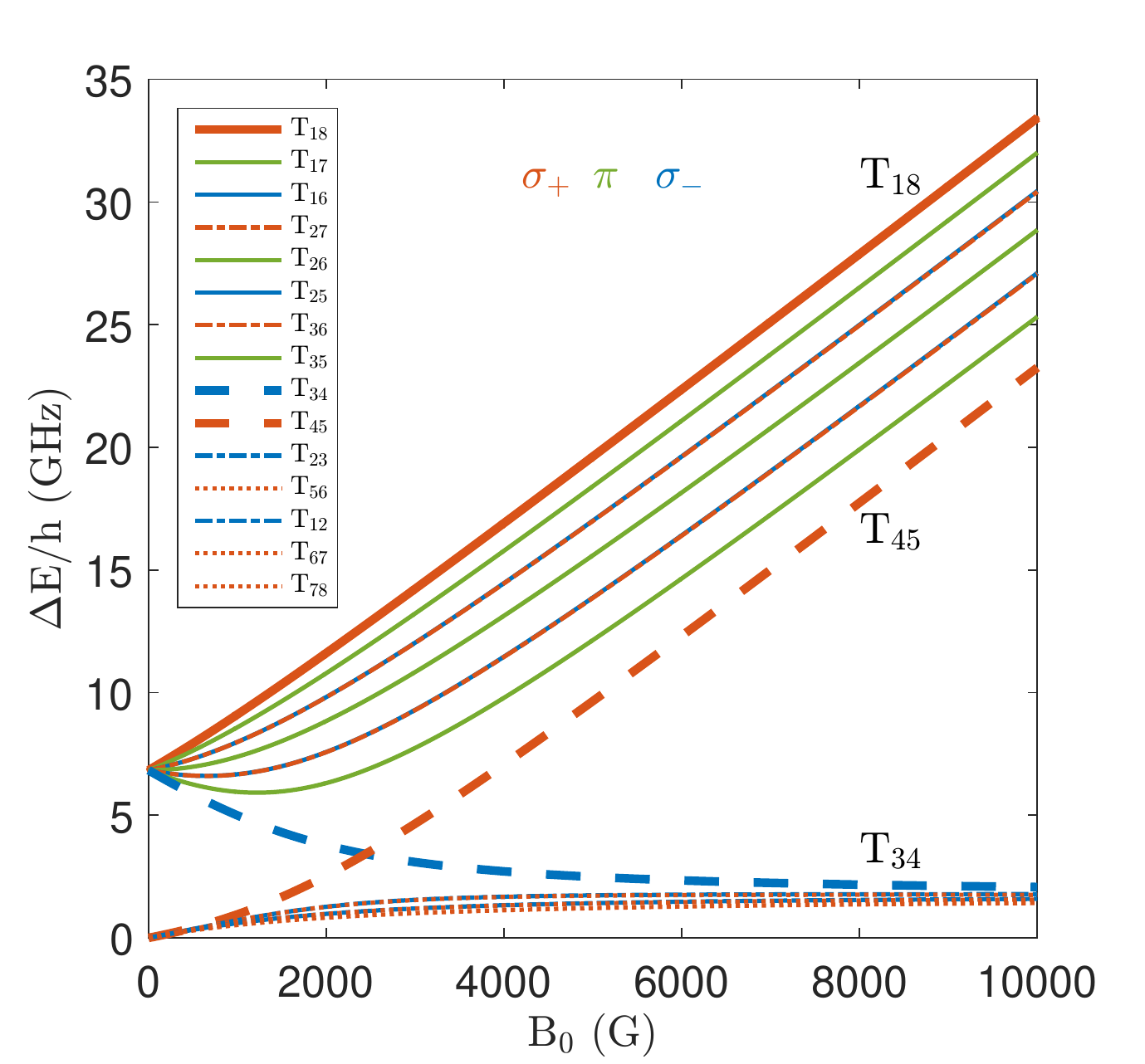}%
\caption{Hyperfine transition frequencies as a function of applied dc magnetic field. The legend lists the transitions in order of decreasing frequency. The $\sigma_+$ transitions are shown in red, $\pi$ transitions in green, and $\sigma_-$ transitions are shown in blue. Dashed lines are used for clarity.}
\label{fig:Rb87_TransitionFrequencies}
\end{figure}

The Hamiltonian for the hyperfine splitting of an atom in an external dc magnetic field \textbf{B}$_{\mathrm{dc}} = B_z \uv{z}$ is
\begin{equation}\label{eq:freqtun_hamiltonian}
H =  H_{\mathrm{hfs}} + H_Z = A_{\mathrm{hfs}} \textbf{I} \cdot \textbf{J} + \mu_B(g_I I_z + g_J J_z) B_z,
\end{equation}
where $H_{\mathrm{hfs}}$ is the hyperfine coupling Hamiltonian, $H_Z$ is the Zeeman Hamiltonian, $\textbf{I}$ and $\textbf{J}$ are the atomic nuclear and electronic spin, respectively, $g_I$ and $g_J$ are the corresponding $g$-factors, and $A_{\mathrm{hfs}}$ is the hyperfine coupling constant~\cite{Steck87}.
The energies of each hyperfine level can be obtained numerically from the eigenvalues of $H$, or analytically by using the Breit-Rabi formula. This is shown as a function of dc magnetic field for the $^{87}$Rb 5S$_{1/2}$ hyperfine levels in Fig.~1 of the main text. The levels are labelled A$_1 \rightarrow \mathrm{A}_8$, in order of increasing energy ($\mathrm{E}_1<\mathrm{E}_2 ... <\mathrm{E}_8$). The resulting hyperfine transition frequencies are shown in Fig.~\ref{fig:Rb87_TransitionFrequencies}. Transitions between levels A$_i$ and A$_f$ are labelled T$_{if}$, with energies E$_{if}=\mathrm{E}_f-\mathrm{E}_i$. The $\sigma_+$ transitions are shown in red, $\pi$ transitions in green, and $\sigma_-$ transitions are shown in blue.

The $H_Z$ term in Eq.~(\ref{eq:freqtun_hamiltonian}) means that as the magnetic field is scanned, the eigenfunctions of the Hamiltonian must change. It is therefore best to describe the hyperfine levels in some field-independent basis, for which the $\ket{I,m_I,J,m_J}$ basis is a convenient choice. As $I=3/2,J=1/2$ for all of the levels, we abbreviate our notation to $\ket{m_I, m_J}$. The composition of each of the hyperfine levels in this basis is given Table~\ref{tbl:level_notation}. In general, each hyperfine level is a superposition of two $\ket{m_I, m_J}$ states, and the coefficients $a$ and $b$ can be determined by numerically diagonalising $H$. The two stretched states, $\ket{F=2,m_F=\pm2} \leftrightarrow \ket{m_I=\pm 3/2, m_J=\pm 1/2}$, are comprised of only a single $\ket{m_I, m_J}$ for all fields. In the absence of dc magnetic field, $a$ and $b$ are given by the Clebsch-Gordon coefficients, listed in Table~\ref{tbl:ab_coefficients_noBfield}. As the dc field strength increases, $a\rightarrow 1$ and $b\rightarrow 0$.


\begin{table}[t!]
\caption{Notation used for the $^{87}$Rb 5S$_{1/2}$ hyperfine levels. The levels A$_1 \rightarrow \mathrm{A}_8$ are in order of increasing energy.}
\centering
\begin{tabular}{c|llll}
 & Weak Field & General  & Strong Field \\
  & ($\ket{F,m_F}$)& ($\ket{m_I,m_J}$) &  ($\ket{m_I,m_J}$) \\
\hline
A$_1$ & $\ket{1,1}$ & $a_1 \ket{3/2,-1/2} + b_1 \ket{1/2,1/2}$ & $\ket{3/2,-1/2}$ \\
A$_2$ & $\ket{1,0}$ & $a_2 \ket{1/2,-1/2} + b_2 \ket{-1/2,1/2}$ & $\ket{1/2,-1/2}$ \\
A$_3$ & $\ket{1,-1}$ & $a_3 \ket{-1/2,-1/2} + b_3 \ket{-3/2,1/2}$ & $\ket{-1/2,-1/2}$ \\
A$_4$ & $\ket{2,-2}$ & $ \ket{-3/2,-1/2}$ & $\ket{-3/2,-1/2}$ \\
A$_5$ & $\ket{2,-1}$ & $a_5 \ket{-3/2,1/2} + b_5 \ket{-1/2,-1/2}$ & $\ket{-3/2,1/2}$ \\
A$_6$ & $\ket{2,0}$ & $a_6 \ket{-1/2,1/2} + b_6 \ket{1/2,-1/2}$ & $\ket{-1/2,1/2}$ \\
A$_7$ & $\ket{2,1}$ &$a_7 \ket{1/2,1/2} + b_7 \ket{3/2,-1/2}$ & $\ket{1/2,1/2}$ \\
A$_8$ & $\ket{2,2}$ & $ \ket{3/2,1/2}$ & $\ket{3/2,1/2}$ \\
\end{tabular}
\label{tbl:level_notation}
\end{table}

\begin{table}[t]
\caption{Coefficients $a$ and $b$ in the $\ket{m_I,m_J}$ basis for the $^{87}$Rb 5S$_{1/2}$ hyperfine levels in the absence of external static magnetic field. The coefficients are the relevant Clebsch-Gordon coefficients for each state. }
\centering
\begin{tabular}{c|lll}
 & $\ket{F,m_F}$& $a$ & $b$ \\
\hline
A$_1$ & $\ket{1,1}$ & $\sqrt{3}/2$ & $-1/2$ \\
A$_2$ & $\ket{1,0}$ &  $1/\sqrt{2}$ & $-1/\sqrt{2}$ \\
A$_3$ & $\ket{1,-1}$ &   $1/2$ & $-\sqrt{3}/2$  \\
A$_4$ & $\ket{2,-2}$ &   &  \\
A$_5$ & $\ket{2,-1}$ &   $1/2$ & $\sqrt{3}/2$ \\
A$_6$ & $\ket{2,0}$ &   $1/\sqrt{2}$ & $1/\sqrt{2}$  \\
A$_7$ & $\ket{2,1}$ &  $\sqrt{3}/2$ & $1/2$ \\
A$_8$ & $\ket{2,2}$ &  &  \\
\end{tabular}
\label{tbl:ab_coefficients_noBfield}
\end{table}

\begin{figure}[t!]
\centering
\includegraphics[width=0.5\textwidth]{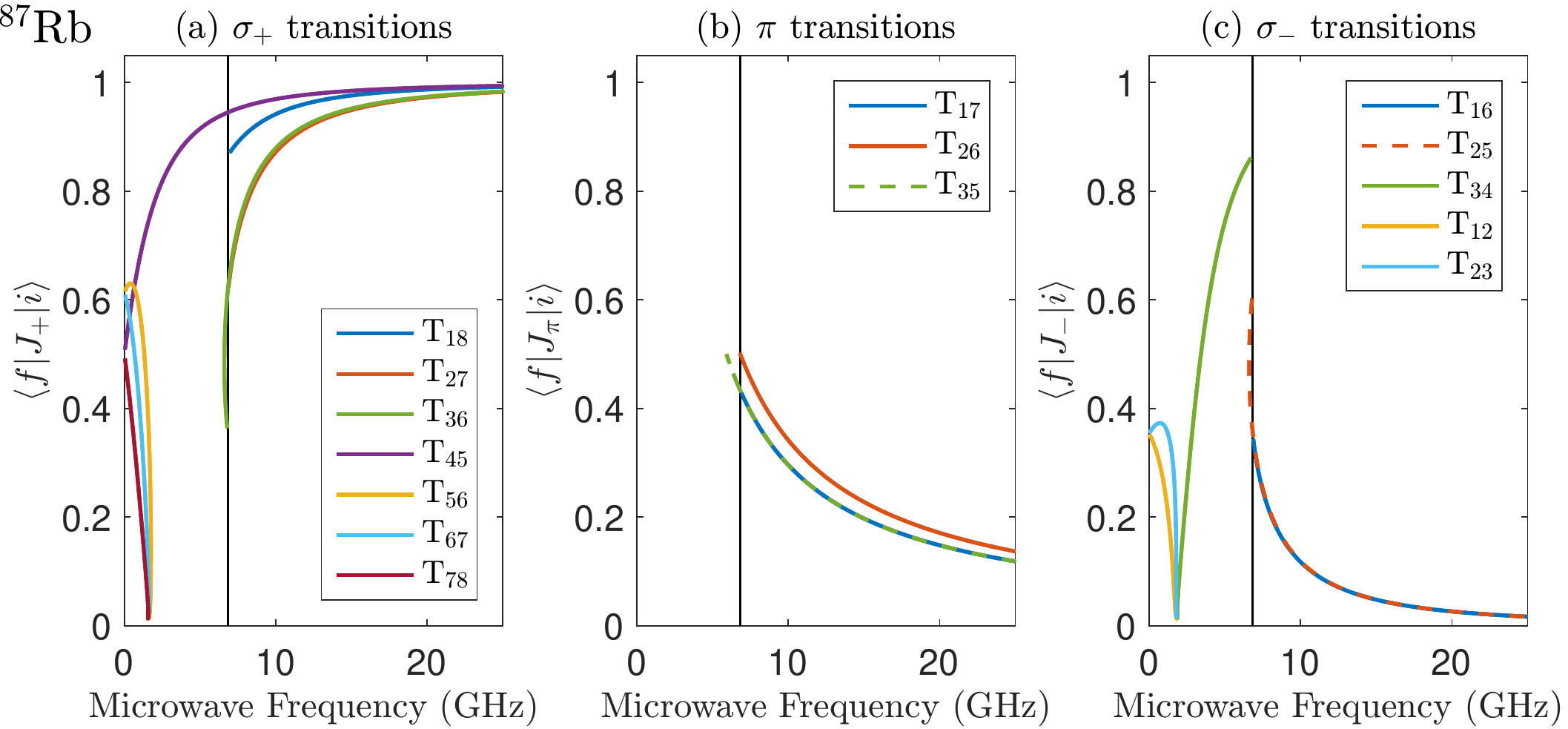}%
\caption{Strengths of the $\sigma_+$, $\pi$, and $\sigma_-$ hyperfine transitions within the $^{87}$Rb $5^2S_{1/2}$ ground state, as a function of the microwave transition frequency. The black vertical line is at 6.835~GHz. Data for $^{87}$Rb was taken from Ref.~\cite{Steck87}.}
\label{fig:transitions_for_microwave_sensing}
\end{figure}

The strengths of the hyperfine transitions also change with the dc magnetic field, and are proportional to the $\bra{f}J_{\gamma}\ket{i}$ matrix element, where $\gamma=-,\pi,+$ is the transition polarisation. We find that $\bra{f}J_{+}\ket{i}=a_ia_f$, $\bra{f}J_{-}\ket{i}=b_ib_f$, and $\bra{f}J_{\pi}\ket{i}=-\tfrac{1}{2}a_ib_f+\tfrac{1}{2}a_fb_i$, meaning that as the magnetic field strength is increased, $\sigma_+$ transitions strengths approach unity, whilst $\sigma_-$ transitions (rapidly) and $\pi$ transitions (slowly) become weak. The $\sigma_+$ transition strengths between states with the same $m_J$ value also go to zero at high dc fields. As a practical comparison, for $\bra{f}J_{\gamma}\ket{i}=1$, a B$_{\mathrm{mw}}=1\,\mu$T microwave field will drive $\Omega_R=2\pi\,28\,\mathrm{kHz}$ Rabi oscillations, whilst for the same microwave field, $\bra{f}J_{\gamma}\ket{i}=0.1$ results in $\Omega_R=2\pi\,2.8\,\mathrm{kHz}$.

%
%

\subsection{Hyperfine Transitions for Microwave Sensing}\label{sec:freqtun_sensing_transitions}

\begin{table*}[t]
\caption{Summary of the transitions available in various alkali species for sensing the $\sigma_+$, $\pi$, and $\sigma_-$ components of a microwave field. $I$ is the nuclear spin. The maximum $\sigma_+$ frequency is defined as the highest transition frequency available at our maximum solenoid field, B$_{\mathrm{dc}}=0.8\,\mathrm{T}$.}
\centering
\begin{tabular}{ccccccc}
Isotope & Abundance & I & E$_{\mathrm{hfs}}/h$ (GHz) & & Min. (GHz) & Max. (GHz) \\
\hline\hline
\multirow{3}{*}{$^{23}$Na}& \multirow{3}{*}{1} & \multirow{3}{*}{$3/2$} & \multirow{3}{*}{1.772} & $\sigma_+$ & 0 & 23.8 \\
 &  &&  & $\pi$ & 1.53 & 5.90 \\
  & & &  & $\sigma_-$ & 0.56 & 2.35 \\
\hline
\multirow{3}{*}{$^{39}$K}& \multirow{3}{*}{0.9326} & \multirow{3}{*}{$3/2$} & \multirow{3}{*}{0.462} & $\sigma_+$ & 0 & 22.8 \\
 &  &&  & $\pi$ & 0.40 & 1.54 \\
  &  &&  & $\sigma_-$ & 0.15 & 0.61 \\
\hline
\multirow{3}{*}{$^{85}$Rb}& \multirow{3}{*}{0.7217} & \multirow{3}{*}{$5/2$} & \multirow{3}{*}{3.036} & $\sigma_+$ & 0 & 25.0 \\
 & &&  & $\pi$ & 2.26 & 10.1 \\
  &  &&  & $\sigma_-$ & 0.74 & 4.15 \\
\hline
\multirow{3}{*}{$^{87}$Rb}& \multirow{3}{*}{0.2783} & \multirow{3}{*}{$3/2$} & \multirow{3}{*}{6.835} & $\sigma_+$ & 0 & 27.9 \\
 &  &  && $\pi$ & 5.92 & 22.8 \\
  &  &&  & $\sigma_-$ & 2.17 & 9.06 \\
\hline
\multirow{3}{*}{$^{133}$Cs}& \multirow{3}{*}{1} & \multirow{3}{*}{$7/2$} & \multirow{3}{*}{9.193} & $\sigma_+$ & 0 & 30.8 \\
 &  &  && $\pi$ & 6.08 & 30.6 \\
  &  & & & $\sigma_-$ & 1.62 & 12.7 \\
\hline
\end{tabular}
\label{tbl:alkali_microwave_transitions}
\end{table*}

The important considerations when choosing a hyperfine transition for microwave sensing are: the microwave frequency of interest, the hyperfine transition strength, the optical resolution of the hyperfine transition states (i.e. the degree to which absorption due to each state can be distinguished), the microwave polarisation of interest, and the dc magnetic field required to tune a hyperfine transition to frequency of interest. Supplementary figure~\ref{fig:transitions_for_microwave_sensing} provides a useful analysis tool, showing the $\sigma_+$, $\pi$, and $\sigma_-$ transition strengths as a function of microwave transition frequency.

T$_{45}$ is the most versatile $\sigma_+$ transition, covering all microwave frequencies above dc. Above 0.6~GHz, it is also the strongest $\sigma_+$ transition for a given microwave frequency. The optical resolution of the neighbouring A$_4$ and A$_5$ states can be poor, however, particularly at low B$_{\mathrm{dc}}$ (corresponding to low microwave frequencies). For microwave sensing of frequencies above 6.835~GHz, the best $\sigma_+$ transition is therefore generally T$_{18}$. The A$_1$ and A$_8$ levels are maximally spectrally resolved from one another, and optical transitions from these levels enjoy minimal background absorption due to $^{85}$Rb. The T$_{18}$ transition is almost as strong as T$_{45}$ for a given microwave frequency, and requires much smaller B$_{\mathrm{dc}}$ to be tuned to a given frequency. For example, to achieve an 18~GHz microwave transition requires B$_{\mathrm{dc}}=0.44\,\mathrm{T}$ on the T$_{18}$ transition, but B$_{\mathrm{dc}}=0.81\,\mathrm{T}$ on the T$_{45}$ transition.

The selection of $\pi$ microwave transition is less clear-cut. T$_{26}$ is the strongest $\pi$ transition, but the difference with T$_{17}$ and T$_{35}$ is not dramatic. T$_{17}$ has the best optical resolution, due to the low $^{85}$Rb absorption background for optical transitions from A$_1$ and the large spectral separation of the A$_1$ and A$_7$ levels. The T$_{35}$ transition is first-order insensitive to dc magnetic fields around $B_{dc}=0.12\,\mathrm{T}$, corresponding to a microwave frequency of 5.92~GHz, and is thus the optimal transition around this point. Supplementary figure~\ref{fig:transitions_for_microwave_sensing} indicates that the $\pi$ transitions can be used for sensing microwaves even above 20~GHz, with the T$_{26}$ transition strength dropping to $\bra{6}J_{\pi}\ket{2}=0.15$ at 22.8~GHz.

The $\sigma_-$ transition strengths quickly drop away for microwave frequencies above 6.835~GHz, with the T$_{16}$ and T$_{25}$ transition strengths dropping to $\bra{f}J_-\ket{i}=0.15$ at 9.06~GHz. However, the T$_{34}$ transition can be used to detect microwaves below 6.835~GHz. The T$_{34}$ transition strength drops to $\bra{4}J_-\ket{3}=0.15$ at 2.17~GHz.

We can perform a similar analysis for other alkali species. The ranges of detectable frequencies for $\sigma_+$, $\pi$, and $\sigma_-$ polarised microwaves using $^{23}$Na, $^{39}$K, $^{85}$Rb, and $^{133}$Cs are summarised in Table~\ref{tbl:alkali_microwave_transitions}. The frequency range was defined as that for which there is a transition with a strength above $\bra{f}J_{\gamma}\ket{i}=0.15$, neglecting transitions between states with the same $m_J$ value. Strong $\sigma_+$ polarised transitions are available at all microwave frequencies and dc magnetic field strengths, and in order to compare the different alkali species, we took the maximum $\sigma_+$ frequency as the highest transition frequency available at our maximum solenoid field, B$_{\mathrm{dc}}=0.8\,\mathrm{T}$. Vapor cells filled with multiple species can be used to span larger frequency ranges. For example, a natural Rb cell provides $\pi$ transitions over the range $2.26-22.8\,\mathrm{GHz}$.


\section{Reconstruction of Microwave Fields of Arbitrary Frequency}

In this section, we provide a framework for reconstructing microwave magnetic fields of arbitrary frequency, using $^{87}$Rb atoms in an applied static magnetic field (\textbf{B}$_{\mathrm{dc}}$) of any strength. This builds on the framework given for the weak dc field regime in Ref.~\cite{Boehi2010a}. The framework is not restricted to $^{87}$Rb, and is valid for microwave transitions in a general system.

In the fixed lab-frame cartesian coordinate system, $(x,y,z)$, a microwave magnetic field is defined by
\[
\textbf{B} \equiv
\begin{pmatrix}
B_{x} e^{-i\phi_{x}}\\
B_{y} e^{-i\phi_{y}} \\
B_{z} e^{-i\phi_{z}} \\
\end{pmatrix}.
\]
In order to reconstruct this field, we need to find these six real values, $B_{x, y, z}$, $\phi_{x, y, z} \in \Re_{\geq 0}$, which are each an implicit function of spatial position. We do this by measuring Rabi oscillations driven by the microwave field on atomic hyperfine transitions. The quantisation axis of our measurements is defined by the applied dc magnetic field, \textbf{B}$_{\mathrm{dc}}$. Following Ref.~\cite{Boehi2010a}, we define a primed cartesian coordinate system, $(x',y',z')$, with the $z'$ axis pointing along the direction of \textbf{B}$_{\mathrm{dc}}$. The $\pi$ and $\sigma_{\pm}$ components of the microwave field in the primed frame are
\begin{equation}
B_- e^{-i\phi_-} \equiv \frac{1}{2}  \Big[ B_{x'} e^{-i\phi_{x'}} + i B_{y'} e^{-i\phi_{y'}}  \Big], \label{eq:B_-} \\
\end{equation}
\begin{equation}
B_{\pi} e^{-i\phi_{\pi}} \equiv B_{z'} e^{-i\phi_{z'}},\label{eq:B_pi} \\
\end{equation}
\begin{equation}
B_+ e^{-i\phi_+} \equiv \frac{1}{2}  \Big[ B_{x'} e^{-i\phi_{x'}} - i B_{y'} e^{-i\phi_{y'}}  \Big],\label{eq:B_+}
\end{equation}
with $B_{-, \pi, +}$, $\phi_{-, \pi, +} \in \Re_{\geq 0}$. For transitions from an initial state $\ket{1}$ to a final state $\ket{2}$, the Rabi frequencies are
\begin{align}
\Omega_- &\equiv \frac{2\mu_B}{\hbar} \bra{2}J_-\ket{1} B_- e^{-i\phi_-}, \label{eq:omega_-} \\
\Omega_{\pi} &\equiv \frac{2\mu_B}{\hbar} \bra{2}J_z\ket{1} B_{\pi} e^{-i\phi_{\pi}}, \label{eq:omega_pi} \\
\Omega_+ &\equiv \frac{2\mu_B}{\hbar} \bra{2}J_{+}\ket{1} B_{+} e^{-i\phi_{+}}, \label{eq:omega_+}
\end{align}
Where $J_z$, $J_+= J_x + iJ_y$, and $J_-= J_x - iJ_y$ are the spin $z$, raising, and lowering operators, respectively.

Note that in the definitions of $B_-$ and $B_+$, a factor of $1/\sqrt{2}$ instead of $1/2$ can also be found in some of the literature. This goes along with a change in the definitions of $J_+$ and $J_-$, which then read $J_{\pm}= \tfrac{1}{\sqrt{2}}(J_x \pm iJ_y)$. If this alternative definition is used, the coefficients $\alpha_+$ and $\alpha_-$ are larger by a factor of $\sqrt{2}$.

\subsection{Microwave Amplitude}
\label{sec:FreqTun_amplitude}

From Eq.~(\ref{eq:omega_pi}), it is straightforward to determine the amplitudes of the microwave magnetic field ($B_{x,y,z}$) when strong $\pi$ transitions are present. The matrix element $\bra{f}J_z\ket{i}$ can be calculated numerically for a any static magnetic field, and so we can obtain the amplitudes along each axis by measuring $\abs{\Omega_\pi}$ with the quantisation axis along $x$, $y$ and $z$ respectively. The $\pi$ (and $\sigma_{-}$) transitions become weak in strong dc fields, however, and in the general case, we need to determine the microwave field amplitudes using only $\sigma_{+}$ transitions.

In the following discussion, the superscript index represents the quantisation axis in the lab frame, i.e. the direction of the applied static magnetic field. Thus for example, $\Omega_+^{+ y}$ ($B_+^{+ y}$) means $\Omega_+$ ($B_+$) for \textbf{B}$_{\mathrm{dc}}$ pointing along the $y$ axis in the positive direction, whilst $\Omega_+^{- y}$ ($B_+^{- y}$) is for \textbf{B}$_{\mathrm{dc}}$ pointing along the $y$ axis in the negative direction.

We begin by finding the sum of $B_-^2$ and $B_+^2$. From Eq.~(\ref{eq:omega_+}) we see that for $\sigma_+$ transitions, we only obtain $B_+$. However, for measurements along a given axis, $B_+$ measured antiparallel to that axis is equivalent to $B_-$ measured parallel to the axis. That is, $B_+^{-} = B_-^{+}$. By measuring $\abs{\Omega_+}$ with the static field both parallel and antiparallel to our axis of measurement, we can thus obtain both $B_+$ and $B_-$. This gives us
\begin{equation}
\label{eq:b-+sum}
B_-^2 + B_+^2 = \frac{\hbar^2}{4\mu_B^2} \Big[ \frac{\abs{\Omega_+^{-}}^2}{\abs{\bra{f}J_+\ket{i}}^2} + \frac{\abs{\Omega_+^{+}}^2}{\abs{\bra{f}J_+\ket{i}}^2} \Big].
\end{equation}
We can also find the $B_-^2 + B_+^2$ sum using Eqs.~(\ref{eq:B_-}) and (\ref{eq:B_+}). Equating this with Eq.~(\ref{eq:b-+sum}) gives
\begin{equation}
\frac{1}{2} (B_{x'}^2 + B_{y'}^2) = \frac{\hbar^2}{4\mu_B^2} \Big[ \frac{\abs{\Omega_+^{-}}^2}{\abs{\bra{f}J_+\ket{i}}^2} + \frac{\abs{\Omega_+^{+}}^2}{\abs{\bra{f}J_+\ket{i}}^2} \Big].
\end{equation}
Defining
\begin{equation}
K_+ \equiv \frac{\abs{\Omega_+^{-}}^2}{\abs{\bra{f}J_+\ket{i}}^2} + \frac{\abs{\Omega_+^{+}}^2}{\abs{\bra{f}J_+\ket{i}}^2},
\end{equation}
we can write
\begin{equation}
\label{eq:B_sum}
B_{x'}^2 + B_{y'}^2 = \frac{\hbar^2}{2\mu_B^2} K_+.
\end{equation}
Next, we apply this formula with the quantisation axis defined along each of the $x$, $y$ and $z$ axes. Starting with \textbf{B}$_{\mathrm{dc}}$ (and thus $z'$) along the $x$ axis, we transform from the primed coordinate system back into the unprimed lab frame, according to the following coordinate transformation:
\begin{eqnarray}
x' &=& -z, \nonumber\\
y' &=& y, \nonumber\\
z' &=& x.
\end{eqnarray}
This transforms the microwave magnetic field phasor to
\[
\textbf{B} \equiv
\begin{pmatrix}
B_{x'} e^{-i\phi_{x'}}\\
B_{y'} e^{-i\phi_{y'}} \\
B_{z'} e^{-i\phi_{z'}} \\
\end{pmatrix} =
\begin{pmatrix}
-B_{z} e^{-i\phi_{z}}\\
B_{y} e^{-i\phi_{y}} \\
B_{x} e^{-i\phi_{x}} \\
\end{pmatrix},
\]
with
\begin{alignat}{2}
B_{x'} &= B_z  &\quad \phi_{x'} &= \phi_z + \pi, \nonumber\\
B_{y'} &= B_y &\quad \phi_{y'} &= \phi_y, \nonumber\\
B_{z'} &= B_x &\quad \phi_{z'} &= \phi_x.
\end{alignat}
Applying this coordinate transformation to Eq.~(\ref{eq:B_sum}) then gives us
\begin{equation}
\label{eq:B_zy}
B_z^2+B_y^2 = \frac{\hbar^2}{4\mu_B^2} K_+^x,
\end{equation}
with $K_+^{x}$ defined in Eq.~(\ref{eq:K+}). We can follow a similar process for \textbf{B}$_{\mathrm{dc}}$ along the $y$ and $z$ axes to get
\begin{equation}
\label{eq:B_zx}
B_z^2+B_x^2 = \frac{\hbar^2}{4\mu_B^2} K_+^y,
\end{equation}
\begin{equation}
\label{eq:B_zy}
B_x^2+B_y^2 = \frac{\hbar^2}{4\mu_B^2} K_+^z.
\end{equation}
Solving these equations simultaneously gives the magnitude of the magnetic field along the (lab frame) $x$, $y$, and $z$ directions,
\begin{align}
B_x^2 &= \frac{\hbar^2}{8\mu_B^2} \Big[ -K_+^x + K_+^y + K_+^z\Big],\label{eq:B_x}\\
B_y^2 &= \frac{\hbar^2}{8\mu_B^2} \Big[ K_+^x - K_+^y + K_+^z\Big],\label{eq:B_y}\\
B_z^2 &= \frac{\hbar^2}{8\mu_B^2} \Big[ K_+^x + K_+^y - K_+^z\Big],\label{eq:B_z}
\end{align}
where $K_+^{\gamma}$ is defined as
\begin{equation}
\label{eq:K+}
K_+^{\gamma} \equiv \frac{\abs{\Omega_+^{-\gamma}}^2}{\abs{\bra{f}J_+\ket{i}}^2} + \frac{\abs{\Omega_+^{+\gamma}}^2}{\abs{\bra{f}J_+\ket{i}}^2}.
\end{equation}
$\abs{\Omega_+^{-\gamma}}$ and $\abs{\Omega_+^{+\gamma}}$ are experimentally determined quantities. The matrix element $\bra{f}J_+\ket{i}$ can be calculated numerically for a general applied static magnetic field, \textbf{B}$_{\mathrm{dc}}$.

\subsection{Microwave Phase}\label{sec:FreqTun_phase}

To reconstruct the field phases, $\phi_x$, $\phi_y$ and $\phi_z$, we begin with the difference of $B_-^2$ and $B_+^2$. Again, this can be found using Eq.~(\ref{eq:omega_+}), measuring $\abs{\Omega_+}$ with the static field both parallel and antiparallel to our axis of measurement, and also using Eqs.~(\ref{eq:B_-}) and (\ref{eq:B_+}):
\begin{align}
B_-^2 - B_+^2 &= \frac{\hbar^2}{4\mu_B^2} \Big[  \frac{\abs{\Omega_+^{-}}^2}{\abs{\bra{f}J_+\ket{i}}^2} - \frac{\abs{\Omega_+^{+}}^2}{\abs{\bra{f}J_+\ket{i}}^2} \Big] \nonumber\\
&= B_{x'} B_{y'} \sin(\phi_{y'}-\phi_{x'}).
\end{align}
This time we define
\begin{equation}
\label{eq:K-}
K_- \equiv \frac{\abs{\Omega_+^-}^2}{\abs{\bra{f}J_+\ket{i}}^2} - \frac{\abs{\Omega_+^+}^2}{\abs{\bra{f}J_+\ket{i}}^2},
\end{equation}
and so we have
\begin{equation}
\label{eq:B_phase}
\sin(\phi_{y'}-\phi_{x'}) =  \frac{\hbar^2}{4\mu_B^2 B_{x'} B_{y'}} K_-.
\end{equation}
We measure $\abs{\Omega_+}^2$ parallel and antiparallel to the $x$, $y$, and $z$ axes, and use Eq.~(\ref{eq:B_phase}) with the same coordinate transformations as in Section~\ref{sec:FreqTun_amplitude}. Inserting the field magnitudes obtained with Eqs.~(\ref{eq:B_x}-\ref{eq:B_z}), we get
\begin{align}
\sin(\phi_{z}-\phi_{y}) &= \label{eq:phi_zy}\\
 2 \Big[ (K_+^x - &K_+^y + K_+^z)  (K_+^x + K_+^y - K_+^z) \Big]^{-1/2} K_-^x, \nonumber\\
\sin(\phi_{x}-\phi_{z}) &=\label{eq:phi_xz}\\
 2 \Big[ ( -K_+^x + &K_+^y + K_+^z)  (K_+^x + K_+^y - K_+^z) \Big]^{-1/2} K_-^y,\nonumber\\
\sin(\phi_{y}-\phi_{x}) &= \label{eq:phi_yx}\\
 2 \Big[ ( -K_+^x + &K_+^y + K_+^z)  (K_+^x - K_+^y + K_+^z) \Big]^{-1/2} K_-^z,\nonumber
\end{align}
with $K_+^{\gamma}$ as defined in equation~\ref{eq:K+} and $K_-^{\gamma}$ defined as $K_-$ for a static magnetic field along the direction $\gamma$:
\begin{equation}
\label{eq:K-_gamma}
K_-^{\gamma} \equiv \frac{\abs{\Omega_+^{-\gamma}}^2}{\abs{\bra{f}J_+\ket{i}}^2} - \frac{\abs{\Omega_+^{+\gamma}}^2}{\abs{\bra{f}J_+\ket{i}}^2}.
\end{equation}
$\abs{\Omega_+^{-\gamma}}$ and $\abs{\Omega_+^{+\gamma}}$ are experimentally determined quantities. The matrix elements $\bra{f}J_-\ket{i}$ and $\bra{f}J_+\ket{i}$ can be calculated numerically for a general applied static magnetic field, \textbf{B}$_{\mathrm{dc}}$. From Eqs.~(\ref{eq:phi_zy}-\ref{eq:phi_yx}), we can obtain the relative phases of the B$_{\mathrm{mw}}$ components.

\bibliography{bibliography_frequency_tunable}